\documentclass[aps,pre,twocolumn,groupedaddress]{revtex4-1}

\usepackage{graphicx} 
\usepackage{epsfig}
\usepackage{amsmath} 
\usepackage{amsthm} 
\usepackage{amssymb}	
\newcommand{\eq}[1]{\begin{align} #1 \end{align}}

\usepackage{graphics} 
\usepackage{hyperref} 
\hypersetup{
    colorlinks,
    citecolor=blue,
    filecolor=black,
    linkcolor=blue,
    urlcolor=blue
}

\newcommand{\abs}[1]{\left| #1 \right|} 
\newcommand{\avg}[1]{\left< #1 \right>} 
\newcommand{\pd}[2]{\frac{\partial #1}{\partial #2}} 
\newcommand{\pdd}[2]{\frac{\partial^2 #1}{\partial #2^2}} 
\newcommand{\ket}[1]{\left| #1 \right>} 
\newcommand{\bra}[1]{\left< #1 \right|} 
\newcommand{\braket}[2]{\left< #1 \vphantom{#2} \right|
 \left. #2 \vphantom{#1} \right>} 
\newcommand{\matrixel}[3]{\left< #1 \vphantom{#2#3} \right|
 #2 \left| #3 \vphantom{#1#2} \right>} 

\theoremstyle{definition}

\usepackage{epstopdf}
\usepackage{epsfig}
\usepackage{tikz}

\begin{document}


\title{The validity of linear response in jammed particulate packings}
\title{Contact nonlinearities and linear response in jammed particulate packings}


\author{Carl P. Goodrich}
\email[]{cpgoodri@sas.upenn.edu}

\author{Andrea J. Liu}
\affiliation{Department of Physics, University of Pennsylvania, Philadelphia, Pennsylvania 19104, USA}

\author{Sidney R. Nagel}
\affiliation{James Franck Institute, The University of Chicago, Chicago, Illinois 60637, USA}

\date{\today}

\begin{abstract}
Packings of frictionless athermal particles that interact only when they overlap experience a jamming transition as a function of packing density.  Such packings provide the foundation for the theory of jamming. This theory rests on the observation that, despite the multitude of disordered configurations, the mechanical response to linear order depends only on the distance to the transition. We investigate the validity and utility of such measurements that invoke the harmonic approximation and show that, despite particles coming in and out of contact, there is a well-defined linear regime in the thermodynamic limit.
\end{abstract}


\maketitle
\section{Introduction}
The harmonic approximation of an energy landscape is the foundation of much of solid state physics~\cite{Ashcroft:1976ud}. Calculations that invoke this simplifying assumption are said to be in the linear regime and are responsible for our understanding of many material properties such as sound propagation and the elastic or vibrational response to small perturbations~\cite{Ashcroft:1976ud,Landau1986}. While the harmonic approximation is not exact and breaks down for large perturbations, the existence of a linear regime is essential to our understanding of ordered solids.

While the lack of any periodic structure has long made amorphous materials difficult to study, the past decade has seen significant progress towards uncovering the origin of commonality in disordered solids by way of the jamming scenario~\cite{Liu:2010jx}. Specifically, numerous studies of the jamming transition of athermal soft spheres have exploited the harmonic approximation to reveal a non-equilibrium phase transition~\cite{Liu:1998up,OHern:2003vq,Liu:2010jx,Wyart:2005wv,Silbert:2005vw,Ellenbroek:2006df,Xu:2007dd,Xu:2010vd,Goodrich:2012ck}. Near this jamming transition, the shape of the landscape near each minimum is essentially the same within the harmonic approximation -- for example, the distribution of curvatures around the minimum, which is directly related to the density of normal modes of vibration, is statistically the same for the vast majority of energy minima.  As a result, linear response properties such as the elastic constants can be characterized by a single property of the minimum, such as its energy, pressure or contact number, which quantifies the distance from the jamming transition for that state.  This powerful property forms the basis of the jamming scenario, which has been shown to explain similarities in the mechanical and thermal properties of many disordered solids. 

However, the jamming scenario is based on systems with finite-ranged potentials. It was pointed out by Schreck {\it et al.}~\cite{Schreck:2011kl} that for such potentials, breaking and forming contacts are a source of nonlinearity and they concluded that the harmonic approximation should not be valid for disordered sphere packings in the large-system limit even for infinitesimal perturbations.  Without a valid linear regime, quantities like the density of vibrational modes and elastic constants are ill defined. Thus, their claim calls into question much of the recent progress that has been made in understanding the nature of the jammed solid.

In this paper, we examine the effect of nonlinearities in jammed sphere packings.  
As we discuss in Sec.~\ref{sec:nonlinearities}, there are two distinct classes of nonlinearities: \emph{expansion nonlinearities} are those that can be understood from the Taylor expansion of the total energy about the local minima, while \emph{contact nonlinearities} are those arising from particles coming in and out of contact. Hentschel {\it et al.}~\cite{Hentschel:2011cba} recently asked whether expansion nonlinearities destroy the linear regime in the thermodynamic limit. By considering carefully the proximity of the system to a plastic rearrangement, which is often preceded by a vibrational mode with vanishing frequency, they concluded that the elastic moduli are indeed well defined.  Here, we provide a detailed analysis of the effect of \emph{contact} nonlinearities; just as Hentschel {\it et al.} found that expansion nonlinearities do not invalidate linear response in the thermodynamic limit, we find that the same is true for contact nonlinearities. Our main results are presented in Sec.~\ref{sec:thermodynamic_limit}, where we show that packings at densities above the jamming transition have a linear regime in the thermodynamic limit despite an extensive number of altered contacts.  We then discuss finite-amplitude vibrations in Sec.~\ref{sec:finite_amplitude_discussion}, and conclude in Sec.~\ref{sec:discussion} with a discussion of our results and their implications for the jamming scenario. 

One somewhat counterintuitive result is that for intrinsically anharmonic potentials such as the Hertzian potential, contact nonlinearities do not affect the harmonic approximation in the limit of small displacements.  Such nonlinearities only pose a danger for Hookian repulsions, but even in that case, there is a well-defined linear regime in the thermodynamic limit for any density above the transition, contrary to the conclusions of Ref.~\cite{Schreck:2011kl}.  Thus, our results show that the harmonic approximation is on footing that is as firm for disordered solids as it is for ordered solids.  

\section{The harmonic approximation and its leading nonlinear corrections\label{sec:nonlinearities}}

We consider athermal packings of $N$ soft spheres in $d$ dimensions that interact with the pair potential
\eq{	
	V_{mn}(r) = \left\{ 
	\begin{array}{l l}
		\frac \epsilon \alpha \left(1-\frac {r}{\sigma}\right)^\alpha & \quad \mbox{if  $r<\sigma$}\\
		0 & \quad \mbox{if $r\geq \sigma$.}\\ 
	\end{array} 
	\right. \label{pair_potential}
}
Here, $r$ is the center-to-center distance between particles $m$ and $n$, $\sigma$ is the sum of the particles' radii, $\epsilon\equiv 1$ sets the energy scale, and $\alpha \ge 2$ determines the power law of the interactions. Such packings jam when the packing fraction $\phi$ exceeds a critical density $\phi_c$, and we will use the excess packing fraction, $\Delta \phi \equiv \phi-\phi_c$, as a measure of the distance to jamming.

The harmonic approximation is obtained from the expansion of the total energy:
\eq{
	U &\equiv \sum_{m,n} V_{mn}(r) \label{total_energy} \\
	&= U^0 - F^0_i u_i + \tfrac 1{2} D^0_{ij}u_i u_j + \tfrac 1{3!}T^0_{ijk}u_iu_ju_k + ... \label{energy_expansion}
}
where the indices $i$, $j$, $k$ run from 1 to $dN$ and index the $d$ coordinates of each of the $N$ particles, and the $dN$-dimensional vector $\vec{u}$ represents some collective displacement about the initial positions. It will be useful to denote the magnitude of $\vec{u}$ as $\delta$ and the direction as $\hat{u}$, so that $\vec{u}=\hat{u}\delta$.
$U^0$ is the energy of the initial system. $\vec F^0$ gives the net force component on every particle, $F^0_i = - \pd{U}{u_i}\big|_{\vec u=0}$, which vanishes if the system is mechanically stable. The dynamical matrix $D^0$ is given by the second derivative of the energy, $D^0_{ij} = \pd{^2U}{u_i \partial u_j}\big|_{\vec u=0}$, and the tensors $T^0$, etc. are given by higher-order derivatives. The ``0" superscripts emphasize that the derivatives are evaluated at $\vec u=0$.

The mechanical response of an athermal system of particles is governed by the equations of motion,
\eq{	m_i \ddot u_i = F_i(\vec u), \label{eq_of_motion}	}
where $m_i$ is the particle mass and $\vec F(\vec u)$ is the vector of instantaneous forces, {\it i.e.}, evaluated at $\vec u$. Since $D_{ij}(\vec u) =- \pd{F_i(\vec u)}{u_j}$, where $D(\vec u)$ is the instantaneous dynamical matrix, this force is generically given by
\eq{	F_i(\vec u) = -\int D_{ij}(\vec u) du_j,	\label{integral_of_dynamical_matrix} }
where the integral follows the trajectory of the particles from the mechanically stable minima at $\vec u = 0$ to the current configuration.

A mechanically stable system is said to be in the linear regime if the harmonic approximation 
\eq{	U-U^0 \approx \tfrac 12 D^0_{ij}u_iu_j	\label{harmonic_approx}}
is accurate enough to describe the phenomenon of interest. Under this assumption, the dynamical matrix is constant and the equations of motion become linear:
\eq{	m_i \ddot u_i = -D^0_{ij}u_j. \label{linearized_eq_of_motion}	}
The solutions to Eq.~\eqref{linearized_eq_of_motion} are called the normal modes of vibration and are among the most studied quantities in solid state physics.

\subsection{Microscopic vs. bulk response}
Importantly, the extent of the linear regime depends on the quantity one wishes to measure; Eq.~\eqref{linearized_eq_of_motion} might accurately describe one phenomena but fail to describe another. Thus, it is important to clarify the quantities of interest~\cite{Goodrich:2014jw}. 
For crystalline solids, the linear approximation is often used to calculate bulk thermal and mechanical properties, such as the elastic moduli and thermal conductivity. However, it is typically \emph{not} used to predict exact microscopic details over long times. If one were to perturb a system along one of its vibrational modes, for example, the linear equations of motion predict simple oscillatory motion confined to the direction of that mode. However, this is not what happens, since even very slight nonlinearities can cause energy to gradually leak into other modes~\cite{Ashcroft:1976ud}. 

Clearly, the linear theory fails to describe such microscopic details, except for the very special case where the harmonic approximation is \emph{exact}, and one would not expect disordered sphere packings to be an exception. However, linear response has had tremendous success in predicting the \emph{bulk} mechanical and thermal properties of crystals. It is these bulk linear quantities, not the details of microscopic response, that are central to the theory of jamming, and will thus be the focus of the remainder of this paper.

We will primarily be concerned with determining whether the harmonic approximation is valid in the limit of infinitesimal displacements, $\delta$.  In other words, we will be asking whether $\delta$ can be made small enough so that Eq.~\eqref{linearized_eq_of_motion} accurately describes bulk response. 
If so, then linear quantities such as the density of states or the elastic constants are well-defined. While experimental measurements in real systems necessarily involve nonzero displacements, our focus on the limit $\delta\rightarrow 0$ will reveal whether the lowest-order behavior can be ascertained from the harmonic approximation.   

To understand the breakdown of the harmonic approximation, it is useful to separate nonlinear corrections into two distinct classes, as outlined below.

\begin{figure}[h!tpb]
	\centering	
	\epsfig{file=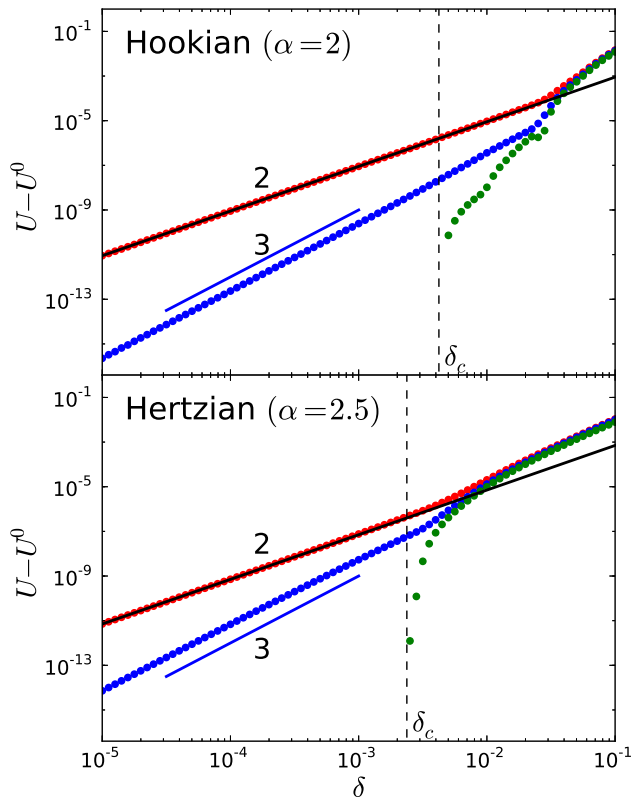,width=1\linewidth,viewport=20 90 480 650, clip}
	\caption{Illustration of nonlinearities for packings of $N=64$ particles with Hookian spring-like interactions ($\alpha=2$, top) and Hertzian interactions ($\alpha=2.5$, bottom). Both systems are at a pressure of $10^{-2}$. The red data shows the total energy $U-U^0$ as the system is displaced by an amount $\delta$ along its lowest non-zero mode. The black line gives the prediction of the harmonic approximation, see Eq.~\eqref{harmonic_approx}. The difference between $U-U^0$ and the predicted energy is shown by the blue data, and the blue line has a slope of $3$. The vertical dashed line represents the value of $\delta$ where the contact network first changes, and the green data gives the magnitude of the change in energy due to contact changes.}
	\label{fig:expansion_nl}
\end{figure}

\subsection{Expansion nonlinearities}
\emph{Expansion nonlinearities} are those which are described by the higher order terms in Eq.~\eqref{energy_expansion}, and can thus be understood from derivatives of the total energy at the energy minimum. However, provided the quadratic term $\tfrac 12 D^0_{ij}u_iu_j$ is positive in all directions, $\delta$ can always be made small enough so that the higher order terms become negligible~\footnote{It is not necessary for $\tfrac 12 D^0_{ij}u_iu_j$ to be positive in the directions of global translation because the energy landscape in these directions is completely flat.}. At the jamming transition, {\it i.e.} $\Delta \phi = 0$, the quadratic term vanishes in some directions in configurational space, so the harmonic approximation fails.  Away from the jamming transition, {\it i.e.} $\Delta \phi>0$, however, the quadratic term is indeed positive in all directions.  (In all our calculations, we remove rattlers, which correspond to zero-frequency modes, so that the dynamical matrix only contains particles that are part of the jammed network.)  Thus, although expansion nonlinearities can be important and even dominate certain phenomena, they cannot prevent a system from having a linear regime provided $\Delta \phi>0$. 

An easy way to observe expansion nonlinearities is to displace a system by an amount $\delta$ in some direction $\hat u$ and measure the energy as a function of $\delta$. $U-U^0$ can then be compared to the prediction of the harmonic approximation given by Eq.~\eqref{harmonic_approx}. A typical example of this is shown in Fig.~\ref{fig:expansion_nl} for jammed packings of particles with Hookian spring-like interactions ($\alpha=2$, top) and Hertzian interactions ($\alpha=2.5$, bottom). 
The corrections to the harmonic approximation have clear cubic behavior at small $\delta$.  Note that they are present when $\alpha=2$: even a spring network has expansion nonlinearities in dimension $d>1$.  This can be seen from Eq.~\eqref{energy_expansion} by writing the tensor $T^0_{ijk}$ as
\eq{
	T^0_{ijk} \equiv{}& \sum_{m,n} \pd{^3V_{mn}(r)}{r_i \partial r_j \partial r_k} \nonumber \\
	={}&\sum_{m,n} t \left(\pd{r}{r_i} \pd{r}{r_j} \pd r {r_k}\right)     - f \left(\pd{^3r}{r_i \partial r_j \partial r_k}\right), \label{d3Vij}     \\
		&+ k \left( \pd{^2r}{r_i \partial r_k} \pd r{r_j} +\pd{^2r}{r_j \partial r_k} \pd r{r_i}  + \pd{^2r}{r_i \partial r_j} \pd r{r_k}\right) \nonumber
}
where $V_{mn}(r)$ is the pair interaction potential of Eq.~\eqref{pair_potential}, $f\equiv -\pd{V_{mn}(r)}{r}$, $k\equiv \pdd{V_{mn}(r)}{r}$, and $t\equiv \pd{^3V_{mn}(r)}{r^3}$, and the terms $\pd r{r_i}$, $\pd{^2r}{r_i \partial r_j}$ and $\pd{^3r}{r_i \partial r_j \partial r_k}$ are generically nonzero.

Clearly, expansion nonlinearities are more dangerous when the harmonic term is small. At the jamming transition ($\Delta \phi=0$), for example, there exist vibrational modes with arbitrarily low frequency that are thus highly susceptible to expansion nonlinearities. Additionally, it is well known that plastic rearrangements in athermal amorphous solids are preceded by a mode frequency going to zero~\footnote{A notable exception is packings of particles with one-sided Hookian interactions.}. Since the density of plastic rearrangements increases with system size, so too does the likelihood that a mode exists with arbitrarily low frequency that is thus highly susceptible to expansion nonlinearities. The effect that this has on the elastic response was studied in detail by Hentschel {\it et al.}~\cite{Hentschel:2011cba}, who showed that the shear modulus is well defined over a small linear regime. For the remainder of this paper, we will focus on the second class on nonlinearities, the contact nonlinearities introduced in Ref.~\cite{Schreck:2011kl}, which we now discuss.

\subsection{Contact nonlinearities\label{sec:contact_nonlinearities}}
Unlike a true spring network, contacts in a sphere packing are allowed to form and break. Since the total energy is a sum over particles in contact, nonlinearities arise when the contact network is altered. Such \emph{contact nonlinearities} cannot be understood from derivatives of the energy at the minimum. For pair interactions of the form of Eq.~\eqref{pair_potential}, the energy expansion of Eq.~\eqref{energy_expansion} is not analytic when contacts form or break and the second derivative is discontinuous if $\alpha\leq 2$. 

For the two systems in Fig.~\ref{fig:expansion_nl}, the green data show the magnitude of the change in energy due to altered contacts.
The vertical dashed lines indicate the minimum displacement magnitude, $\delta_c$, required to change the contact network.  While the value of $\delta_c$ varies greatly depending on the realization and displacement direction, Schreck {\it et al.}~\cite{Schreck:2011kl} showed that $\avg{\delta_c}\rightarrow 0$ in two important limits. As the number of particles increases, so too does the number of contacts and thus the probability that \emph{some} contact is on the verge of forming or breaking must also increase. Similarly, all contacts are on the verge of breaking in a marginally jammed system at $\Delta \phi = 0$. Therefore, the onset amplitude $\delta_c$ of contact nonlinearities vanishes as either $N\rightarrow \infty$ or $\Delta \phi \rightarrow 0$~\cite{Schreck:2011kl}.

\subsection{Important limits}
Due to the existence of a phase transition at the jamming point, the limits $N\rightarrow \infty$ and $\Delta\phi \rightarrow 0$ are of particular interest. When studying the leading order mechanical properties of a solid, one also considers the limit of infinitesimal displacements, {\it i.e.} $\delta \rightarrow 0$. However, the order at which these limits are taken is important. For example, Schreck {\it et al.} showed that $\delta_c>0$ for finite $\Delta \phi$ and $N$~\cite{Schreck:2011kl}, so there is a perfectly well-defined linear regime if $\delta\rightarrow 0$ is the first limit taken.  This is the standard order of limits taken, for example, in the harmonic theory of crystalline solids~\cite{Ashcroft1976}.  

We already saw that expansion nonlinearities can occur if one considers taking the limit $\Delta \phi \rightarrow 0$ before $\delta \rightarrow 0$, and the importance of these nonlinearities is emphasized in, for example, Refs.~\cite{OHern:2003vq,Xu:2010fa,Gomez:2012ji,Ikeda:2013gu}. Furthermore, Schreck {\it et al.}~\cite{Schreck:2011kl,Schreck:2013gg} showed that contact nonlinearities will also be present in this case, regardless of system size. 
Thus, there is no linear regime at $\Delta \phi=0$. This result was generalized to finite temperatures by Ikeda {\it et al.}~\cite{Ikeda:2013gu} and Wang and Xu~\cite{Wang:2013gy}, who independently showed that the linear regime breaks down above a temperature $T^*$ when $\Delta \phi > 0$.

Finally, for athermal systems above the jamming transition ($\Delta \phi>0$), contact nonlinearities are unavoidable if  we take the limit $N\rightarrow \infty$ before $\delta \rightarrow 0$.   Nevertheless, we will show next that there is still a well-defined linear regime in this case.

\section{The linear regime in the thermodynamic limit\label{sec:thermodynamic_limit}}
In this section, we will show that there is always a well-defined linear regime in the thermodynamic limit whenever $\Delta \phi>0$. We will assume that $T=0$ and that the limit $N\rightarrow \infty$ is taken before the limit $\delta \rightarrow 0$ so that any infinitesimal displacement $\delta\ket{\hat u}$ changes the contact network, leading to contact nonlinearities. 
As discussed above, we will primarily be concerned with establishing the existence of a linear regime for bulk quantities, such as the elastic constants or heat capacity. Since these quantities are described by the density of vibrational modes, $D(\omega)$, it will suffice to show that $D(\omega)$ is insensitive to nonlinear corrections in the limit $\delta \rightarrow 0$. This is not the case for microscopic quantities, such as the precise time evolution following a particular perturbation to a particle or group of particles, which can be highly sensitive to microscopic details that have no noticeable bulk effect.

We will first present a perturbation argument to show that changes to $D(\omega)$ due to contact nonlinearities vanish in the thermodynamic limit as $N^{-1}$~\cite{Goodrich:2014jw}. This result is independent of potential and shows that linear response holds for bulk quantities. We will then present a far simpler argument, based on the continuity of the dynamical matrix for potentials with $\alpha >2$, that shows a clear linear regime for \emph{both} bulk and microscopic quantities~\cite{Goodrich:2014jw}. Our results can be reconciled with those of Schreck {\it et al.}~\cite{Schreck:2011kl,Schreck:2013gg} because they only look at microscopic quantities of relatively small packings close to the transition.

\subsection{Validity of bulk linear response\label{sec:perturbation_argument}}
Here, we will construct a perturbation theory to describe the effect of contact nonlinearities on the vibrational modes and their corresponding frequencies.  We will begin by considering only a single altered contact and then extend the results to the case of many altered contacts. We will assume that $N^{-1}\ll \delta\ll 1$ so that contact nonlinearities are unavoidable but all expansion nonlinearities can be ignored. 

Let $\Delta D$ be the change in the dynamical matrix as a result of the change of a single contact, so that the new dynamical matrix is $\tilde D = D^0 + \Delta D$. Note that $\Delta D$ is highly sparse with only $4d^2$ non-zero elements, where $d$ is the dimensionality. We now consider the effect of the perturbation $\Delta D$ on the eigenmodes of $D^0$ ({\it i.e.}, the original normal modes of vibration).

\subsubsection{Extended modes}
Let $\ket{\hat e_n}$ and $\omega_n^2$ be the $n$th eigenmode and eigenvalue of $D^0$, respectively. If a normalized mode is extended, then every component will be of order $N^{-1/2}$.  For now, we will assume that all modes are extended; the extension of the argument to localized modes is discussed below. The change in the $n$th eigenvalue of $D^0$ can be described by the expansion
\eq{	
	\Delta \omega_n^2 & \equiv \tilde \omega_n^2 - \omega_n^2 \nonumber \\
	&= \matrixel{\hat e_n}{\Delta D}{\hat e_n} + \sum_{m \neq n} \frac{\abs{\matrixel{\hat e_m}{\Delta D}{\hat e_n}}^2}{\omega_n^2 - \omega_m^2} + ...\label{eigenvalue_perturbation_2order}
}
where $\tilde \omega_n^2$ is the eigenvalue of $\tilde D$ and $\matrixel{\hat e_n}{\Delta D}{\hat e_n} \sim \matrixel{\hat e_m}{\Delta D}{\hat e_n} \sim N^{-1}$ because the modes are extended and $\Delta D$ is highly sparse. While the first-order term clearly scales as $N^{-1}$, the higher-order terms depend on the mode spacing as well. Since the probability distribution of eigenvalues does not depend on $N$, the average eigenvalue spacing is proportional to $N^{-1}$. If we assume that 
\eq{	\abs{\omega_n^2-\omega_m^2} > N^{-1}, \label{nondegenerate_criteria}}
then all higher order terms in Eq.~\eqref{eigenvalue_perturbation_2order} are at most proportional to $N^{-1}$. 

However, just because the average mode spacing is of order $N^{-1}$ does not mean that \emph{all} modes are separated by $N^{-1}$. To account for the possibility of, for example, two nearly degenerate modes, $\ket{\hat e_n}$ and $\ket{\hat e_m}$, that do not satisfy Eq.~\eqref{nondegenerate_criteria}, we explicitly solve the degenerate perturbation problem given by
\eq{	
	V \equiv \left( 
	\begin{array}{c c}
		\matrixel{\hat e_n}{\Delta D}{\hat e_n} &\matrixel{\hat e_m}{\Delta D}{\hat e_n} \\
		\matrixel{\hat e_n}{\Delta D}{\hat e_m} &\matrixel{\hat e_m}{\Delta D}{\hat e_m}
	\end{array}
	\right)
	\label{V_extended_only}
}
that treats the coupling between $\ket{\hat e_n}$ and $\ket{\hat e_m}$.
The eigenvalues of $V$ give the full corrections to $\omega_n^2$ and $\omega_m^2$ from their mutual interaction with the perturbation $\Delta D$. The coupling with the other modes is given by Eq.~\eqref{eigenvalue_perturbation_2order}, where terms involving the two nearly degenerate modes are omitted. 

Since the elements of $V$ are all proportional to $N^{-1}$, so too are its eigenvalues. We have already shown that the non-degenerate effect is at most order $N^{-1}$, so the full effect of the perturbation on all eigenvalues $\omega_n^2$ must vanish in the thermodynamic limit. 

We can construct a similar expansion for the eigenvectors. The non-degenerate case is given by
\eq{	\ket{\tilde{e}_n} = \ket{\hat e_n} + \sum_{m\neq n} \frac{\matrixel{\hat e_m}{\Delta D}{\hat e_n}}{\omega_n^2 - \omega_m^2} \ket{\hat e_m} + ...	 \label{eigenvector_expansion}}
while the coupling between nearly degenerate modes that do not satisfy Eq.~\eqref{nondegenerate_criteria} is given by the eigenvectors of $V$. As should be expected, the eigenvectors of $V$ can cause considerable mixing between the nearly degenerate modes. 
Furthermore, the coefficients in front of $\ket{\hat e_m}$ in Eq.~\eqref{eigenvector_expansion} do not vanish when $\abs{\omega_n^2-\omega_m^2}$ is of order $N^{-1}$. Thus, an eigenmode can mix with the few modes nearest in frequency, but the eigenvalue difference between such modes vanishes as $N^{-1}$. In the thermodynamic limit, modes that are able to mix must already be degenerate, so distinguishing between them is meaningless. It is clear that the mode mixing caused by the perturbation $\Delta D$ cannot change the spectral density in the thermodynamic limit.

\subsubsection{Localized modes}
We will now consider the effect of localized modes. We will show that localized modes that overlap with the altered contact can change substantially, but their presence does not affect the extended modes. Furthermore, since the number of modes that are localized to a given region cannot be extensive, the total density of states will be unaffected. Although we will consider modes that are completely localized to a few particles, the arguments can be easily applied to quasi-localized modes by including appropriate higher-order corrections. 

If a localized mode does not overlap with the altered contact, then the matrix elements in Eqs.~\eqref{eigenvalue_perturbation_2order} and \eqref{eigenvector_expansion} involving that mode are zero. In this trivial case, the localized mode is unchanged and does not couple to any other modes. However, if a localized mode \emph{does} overlap with the altered contact, then the matrix elements coupling it to an extended mode are proportional to $N^{-1/2}$ (not $N^{-1}$, as it is for the extended modes).

In this case, we cannot use the non-degenerate perturbation theory of Eqs.~\eqref{eigenvalue_perturbation_2order} and \eqref{eigenvector_expansion}. For instance, there is always a $k$th order term in Eq.~\eqref{eigenvalue_perturbation_2order} that is proportional to $N^{-1}/\abs{\omega_n^2-\omega_m^2}^{k-1}$ and does not converge unless $\abs{\omega_n^2-\omega_m^2} \gg \mathcal{O}\left(N^{-1/(k-1)}\right)$. Therefore, we must treat the interaction between the localized mode and all nearby extended modes by solving the degenerate problem. 

Let $\omega_l^2$ be the the eigenvalue of a localized mode, and let the indices $s$ and $t$ run over the set of $\rho N$ modes that satisfy
\eq{	\abs{\omega_{s,t}^2 - \omega_l^2} < c \nonumber}
where $c$ is some small constant. Note that the localized mode is among those spanned by $s$ and $t$.
To diagonalize the symmetric perturbation matrix
\eq{ V_{st} \equiv \matrixel{\hat e_s}{\Delta D}{\hat e_t},}
note that the dynamical matrix can be written~\cite{Pellegrino:1993uu} as
\eq{ D = AF^{-1}A^T.  \nonumber}
Here, $A$ is the equilibrium matrix and has $dN$ rows and $N_c$ columns, where $N_c$ is the number of contacts, $N$ is the number of particles and $d$ is the dimensionality. $F$ is the diagonal flexibility matrix and has $N_c$ elements $F_{ii} = 1/k_i$, where $k_i$ is the stiffness of the $i$th contact. When $N_c=1$, as is the case for our perturbation matrix $\Delta D$, the equilibrium matrix becomes a vector, $A\rightarrow\ket{A}$, and the flexibility matrix becomes the scalar $1/k$. We can now write the matrix elements as
\eq{
	V_{st} &= \matrixel{\hat e_s}{\Big(\ket{A}k \bra{A}\Big)}{\hat e_t} \nonumber \\
	&= a_s a_t,
	\label{V_reduced}
}
where $a_s = k^{1/2}\braket{A}{\hat e_s}$ and $\braket{A}{\hat e_s}$ is simply the projection of the original eigenvector $\ket{\hat e_s}$ onto the broken contact. Note that for extended modes, the magnitude of $a_s$ scales as
\eq{	a_s  \sim N^{-1/2} 	}
while for localized modes
\eq{	a_l  \sim 1.}
The eigenvalues and eigenvectors of the $\rho N$ by $\rho N$ matrix $V_{st}$ can be solved exactly, with the following results.

A matrix of the form of Eq.~\eqref{V_reduced} has only one non-zero eigenvalue,
\eq{	\Delta \omega_l^2 = a_l^2 + \sum_{s\neq l} a_s^2,	}
which gives the change in energy of the localized mode and does not vanish in the thermodynamic limit. This is not surprising given the drastic overlap between the mode and the altered contact. Similarly, the corresponding eigenvector gives the coupling from the extended modes:
\eq{	\ket{\tilde e_l} = \ket{\hat e_l} + \sum_{s\neq l} \frac{a_s}{a_l}\ket{\hat e_s}.	}
From the scalings of $a_s$ and $a_l$, there is a $N^{-1/2}$ contribution to $\ket{\tilde e_l}$ from each of the $\rho N$ extended modes, the elements of which also scale like $N^{-1/2}$. Therefore, $\ket{\tilde e_l}$ becomes at least partially extended if $\sum_{s\neq l} a_s^2 > 0$.

Thus, forming or breaking a single contact can significantly change the eigenvalue of localized modes that happen to overlap with the altered contact. However, since the density of such modes vanishes as $N^{-1}$, the effect on the density of states in negligible. Furthermore, note that if the initial displacement $\ket{u}$ is along a localized mode, then there is always a finite displacement amplitude $\delta_c$ before the first contact change and so contact nonlinearities can be avoided. 

To understand the effect of localized modes on the extended modes, we see that all other eigenvalues of $V_{st}$ are zero:
\eq{	\Delta \omega_s^2 = 0 \;\;\;\; \text{for all } s\neq l	}
This implies that the frequency of an extended mode does not change due to the presence of a localized mode. However, there is a small correction to the mode itself,
\eq{	\ket{\tilde e_s} &= \ket{\hat e_s} - \frac {a_s}{a_l} \ket{\hat e_l}, \;\;\;\; \text{for all } s\neq l	}
but this correction vanishes in the thermodynamic limit.

Thus, we have shown that even for Hookian springs, altering a single contact in the thermodynamic limit cannot change the density of states~\cite{Goodrich:2014jw}. For extended modes, eigenvalues can change by at most order $N^{-1}$ and mode mixing is allowed only between modes whose eigenvalue spacing is less than $N^{-1}$. While localized modes that overlap with the altered contact can have a non-negligible change in eigenvalue and mix with a large number of extended modes, the density of such localized states vanishes as $N^{-1}$. 

So far we have considered the effect of changing a single contact. We can find an upper bound for the total number of contacts that can change, $\Delta N_c$, by considering the distribution $P(r - \sigma)$, where $r$ is the center-to-center distance between two particles and $\sigma$ is the sum of their radii. $P(x)$ measures the likelihood that two particles are a distance $x$ away from \emph{just} touching, and is conceptually very similar to the radial distribution function (the two are identical for monodisperse packings). Since a contact can only change if $\left| r-\sigma\right| \lesssim \delta$, where $\delta$ is again the perturbation amplitude, we can approximate $\Delta N_c$ by integrating $P(x)$ from $-\delta$ to $\delta$ and multiplying by the system size. For finite $\Delta \phi$, $P(r-\sigma)$ is finite at $r=\sigma$, so the total number of altered contacts is
\eq{	\Delta N_c \sim N \delta  \qquad \text{ for $\Delta \phi > 0$}. \label{eq:DeltaNc}}

Although this diverges when the limit $N\rightarrow \infty$ is taken before $\delta \rightarrow 0$, the density of altered contacts $\Delta N_c/N$ vanishes. Using the above result that each altered contact affects the density of states by at most $N^{-1}$, we see that the net effect of altering $\Delta N_c$ contacts is proportional to $\delta$. Thus, even when an extensive number of contacts are altered in the thermodynamic limit, the effect on the density of states vanishes as $\delta \rightarrow 0$ and we conclude that the linear regime is well defined.

Finally, we can use this result to estimate how the size of the linear regime vanishes in the limit $\Delta \phi \rightarrow 0$. Here, we will only consider the effect of contact nonlinearities; see the supplementary material for a rough estimation of when expansion nonlinearities become important. 
Like the radial distribution function, $P(r-\sigma)$ forms a $\delta$ function at $r=\sigma$ when $\Delta\phi=0$. This means that even for arbitrarily small perturbations, a macroscopic number of contacts change, implying that the above argument does not hold when the limit $\Delta \phi \rightarrow 0$ is taken before $\delta \rightarrow 0$. 
For small but finite $\Delta \phi$, the peak in $P(r-\sigma)$ shifts slightly and its height is proportional to $\Delta \phi^{-1}$~\cite{OHern:2003vq}. Therefore, we can overestimate the above integral by assuming $P(r-\sigma) \sim \Delta\phi^{-1}$ over the range of integration. Equation~\eqref{eq:DeltaNc} becomes $\Delta N_c \sim N \delta / \Delta \phi$, so the net effect of contact nonlinearities on the density of states is proportional to $\delta / \Delta \phi$. Setting this to a constant, determined by the ``acceptable" amount of deviation from linear behavior, we see that the displacement amplitude at which contact nonlinearities become important vanishes linearly with $\Delta\phi$.

\begin{figure}
	\centering	
	\epsfig{file=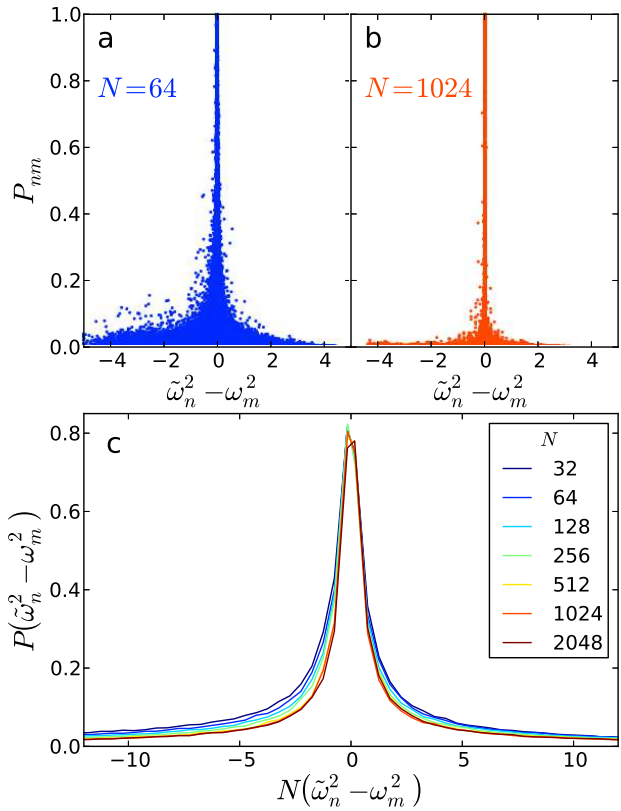,width=1\linewidth,viewport=20 80 480 655, clip}
	\caption{The projection of a perturbed mode $\ket{\tilde e_n}$ onto the original modes $\ket{\hat e_m}$ as a function of the eigenvalue difference. a) 16 realizations of $N=64$ particle systems. b) 1 realization of a $N=1024$ particle system. All systems have Hookian interactions ($\alpha=2$) in 2 dimensions and are at a pressure of $10^{-2}$. c) The average projection as a function of $N\left(\tilde \omega_n^2-\omega_m^2\right)$ for various system sizes. The range of $\tilde \omega_n^2-\omega_m^2$ over which the projection is relevant vanishes slightly faster than $N^{-1}$.}
	\label{fig:mode_projection}
\end{figure}

\subsubsection{Numerical Verification}
We now provide numerical evidence to support the analytical result that changing a single contact has a $N^{-1}$ effect on the linear vibrational properties. To do this, we generate mechanically stable 2-dimensional packings of spheres that interact according to Eq.~\eqref{pair_potential} with $\alpha=2$. For each mechanically stable system, we first obtain the normal modes of vibration by diagonalizing the dynamical matrix $D^0$. We then perturb the system by removing the weakest contact without actually displacing any particles. This perturbed system no longer corresponds to a sphere packing but allows us to isolate contact nonlinearities without considering expansion nonlinearities. The diagonalization of the resulting dynamical matrix $\tilde D$ gives the normal modes of vibration for the perturbed system.

We compare the vibrational modes in Fig.~\ref{fig:mode_projection} by projecting each mode $\ket{\tilde e_n}$ of the perturbed system onto each mode $\ket{\hat e_m}$ of the unperturbed system. The projection 
\eq{ P_{nm} \equiv \braket{\tilde e_n}{\hat e_m} }
quantifies how close a perturbed mode is to an unperturbed mode. Fig.~\ref{fig:mode_projection}a shows a scatter plot of $P_{nm}$ as a function of the difference in eigenvalue $\tilde \omega_n^2 - \omega_m^2$ for 16 systems of $N=64$ particles at a pressure of $10^{-2}$. Fig.~\ref{fig:mode_projection}b shows similar data but for a system of $N=1024$ particles. As expected, the projection has a sharp peak at $\tilde \omega_n = \omega_m$, because mode mixing is stronger among modes of the same frequency. 

The width of the peak in the projection is clearly smaller for the larger system. The $N$-dependence of the width is quantified in Fig.~\ref{fig:mode_projection}c, which shows the average projection, $P\left(\tilde{\omega}_n^2 - \omega_m^2\right)$, as a function of $N\left(\tilde{\omega}_n^2 - \omega_m^2\right)$. By comparing the width of $P\left(\tilde{\omega}_n^2 - \omega_m^2\right)$ at different system sizes, we see that it vanishes slightly faster than $N^{-1}$, confirming the fundamental result of our perturbation calculation above. 

\begin{figure}
	\centering	
	\epsfig{file=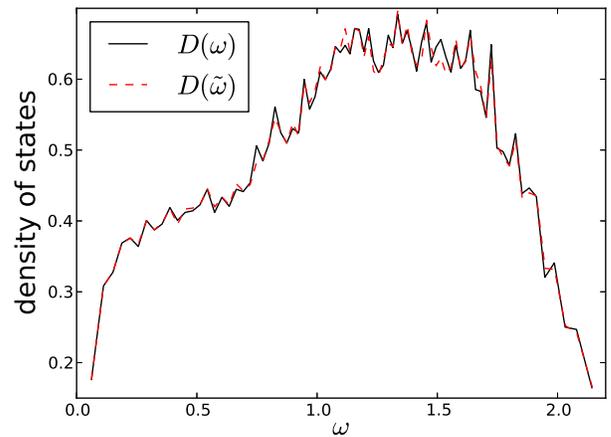,width=1\linewidth,viewport=0 0 480 330, clip}
	\caption{Density of states for a single mechanically stable system ($D(\omega)$, solid black curve) and an identical system with a single contact removed ($D(\tilde \omega)$, dashed red curve). The $N=2048$ particle system has Hookian interactions and is at a pressure of $10^{-2}$. The difference between the two density of states is well within the inherent fluctuations.}
	\label{fig:dos_change}
\end{figure}

We can also measure the shift in vibrational frequency due to the removal of a single contact. The solid black curve in Fig.~\ref{fig:dos_change} shows the density of vibrational states, $D(\omega)$, for a single mechanically stable $N=2048$ particle system. The dashed red curve shows the density of states, $D(\tilde \omega)$, for the corresponding perturbed system. While there is a small change, this difference is not systematic and is much smaller than the fluctuations inherent in the measurement. Since the density of states remains non-zero down to a characteristic frequency $\omega^*$, which is related to the number of contacts above jamming, $\Delta Z$~\cite{Wyart:2005wv,Silbert:2005vw}, changes to the contact network could have a more drastic effect if $\Delta Z \lesssim N^{-1}$. In the thermodynamic limit, however, this only occurs when $\Delta \phi \sim \Delta Z^{2} \rightarrow 0$, {\it i.e.}, at the jamming transition, where nonlinear effects are known to dominate.

\subsection{Continuity of the dynamical matrix for $\alpha>2$}
In the above perturbation argument, we looked at the effect of forming or breaking a contact on the eigenmodes and frequencies of the system. We exploited the sparsity of the perturbation matrix $\Delta D$ to show that the effects scale like $N^{-1}$, but we allowed the non-zero values of $\Delta D$ to be finite in magnitude. If these non-zero elements were vanishingly small, however, then the effect of altering the contact would be negligible and the above perturbation argument would not be necessary. 
We will see that this is the case for potentials where the dynamical matrix is a continuous function of the particle positions.  In that case, the forming or breaking of a contact has a negligible effect on the response in the limit of small displacements. 

We begin by considering the difference between the linear and nonlinear equations of motion (Eqs.~\eqref{eq_of_motion} and \eqref{linearized_eq_of_motion}). If we define $F^\text{harm}_i(\vec u)\equiv -D^0_{ij}u_j$, then the quantity
\eq{	\Delta F_i(\vec u) \equiv \left|F_i^\text{harm}(\vec u)- F_i(\vec u)\right|/\left|F_i^\text{harm}(\vec u)\right|	}
measures the relative error associated with the linearized equations of motion. If $\Delta F_i(\vec u)$ remains ``suitably" small, which again depends on the quantity being measured, then the harmonic approximation is justified and there is a valid linear regime.

Note that $\left|F_i^\text{harm}(\vec u)\right|$ is clearly proportional to $\delta$. Furthermore, using Eq.~\eqref{integral_of_dynamical_matrix} we can write $F_i^\text{harm}(\vec u)- F_i(\vec u) = \int  \left( D_{ij}(\vec u)-D_{ij}^0\right) du_j$, where $D_{ij}(\vec u)$ are the elements of the instantaneous dynamical matrix at displacement $\vec u$. If all elements $ D_{ij}(\vec u)-D_{ij}^0$ vanish in the limit $\delta \rightarrow 0$, then $F_i^\text{harm}(\vec u)- F_i(\vec u)$ must vanish \emph{faster} than $\delta$ and $\Delta F_i(\vec u) \rightarrow 0$ as $\delta \rightarrow 0$.

Also note that the instantaneous dynamical matrix can generically be written as
\eq{
	D_{ij}(\vec u) \equiv{}& \sum_{\text{contacts}} \pd{^2V(r)}{r_i \partial r_j} \nonumber \\
	={}&\sum_{\text{contacts}} k(r) \pd{r}{r_i} \pd{r}{r_j} -f(r) \pd{^2r}{r_i \partial r_j}, \label{D_ij}
}
where $f\equiv -\pd{V(r)}{r}$ and $k\equiv \pdd{V(r)}{r}$ are the force and stiffness of each contact, respectively, evaluated at $\vec u$. Therefore, we see that if $f(r)$ and $k(r)$ are continuous functions of the distance $r$ between two particles, then $D_{ij}(\vec u)$ is a continuous function of particle positions, which implies that $D_{ij}(\vec u)-D_{ij}^0$ vanishes for small $\delta$ and there is a valid linear regime.

Now, for one-sided interaction potentials of the form of Eq.~\eqref{pair_potential}, $f(r)$ is given by  
\eq{
	f(r) & \equiv \pd{V(r)}{r} = \left\{
		\begin{array}{l l}
			\frac{\epsilon}{\sigma} \left(1-\frac r\sigma\right)^{\alpha-1} & \quad \mbox{if $r<\sigma$}\\
			0 & \quad \mbox{if $r \geq \sigma$}
		\end{array} \right. 
}
and $k(r)$ is given by 
\eq{
	k(r) & \equiv \pdd{V(r)}{r} \nonumber \\
	&= \left\{
		\begin{array}{l l}\frac{\epsilon(\alpha-1)}{\sigma^2} \left(1-\frac r\sigma\right)^{\alpha-2} & \quad \mbox{if $r<\sigma$}\\
		0 & \quad \mbox{if $r \geq \sigma$}
		\end{array} \right. .
	\label{k_of_r}
}
$f(r)$ and $k(r)$ are both continuous when $r<\sigma$ and when $r>\sigma$; it is the point of contact ($r=\sigma$) that poses a potential problem. Discontinuities do indeed arise when the exponent $\alpha$ is less than or equal to 2, but $f(r)$ and $k(r)$, and thus $D_{ij}(\vec u)$, are clearly continuous whenever $\alpha > 2$. Thus, there is always a valid linear regime for interaction potentials with $\alpha>2$~\cite{Goodrich:2014jw}.

We can calculate a lower bound for the size of the linear regime by requiring that the change in any element of $D_{ij}(\vec u)$ never exceeds some $\Delta D_\text{max}$. This is satisfied if the change in $k(r)$ of any bond never exceeds $\Delta D_\text{max}$. From Eq.~\eqref{k_of_r}, we see that the maximum change in contact length, $\Delta r_\text{max}$, is given by
\eq{	\frac{\Delta r_\text{max}}{\sigma} = \left( \frac{\Delta D_\text{max} \sigma^2}{\epsilon \left(\alpha-1\right)}\right)^{1/(\alpha-2)}. \label{Deltar_max}	}
Therefore, if $\delta \braket{\hat u}{\!r}$ is the projection of the displacement onto the bond length $r$, then the system has a well-defined linear regime for $\delta < \delta_0$~\cite{Goodrich:2014jw}, where
\eq{	\delta_0 = \Delta r_\text{max} / \braket{\hat u}{\!r}.}

This statement is valid for any potential $\alpha>2$ and is independent of the number of contacts that change or the system size. Importantly, the limit $\alpha \rightarrow 2^+$ is still well-behaved so that it is only in the case $\alpha=2$ that $\delta_0=0$ and we must resort to the perturbation argument presented above in Sec.~\ref{sec:thermodynamic_limit}. 

In Ref.~\cite{Schreck:2013gg}, Schreck {\it et al.} showed that for systems with Hertzian interactions, contact nonlinearities have a smooth effect on the spectral density as $\delta$ increases above $\delta_c$, the minimum displacement magnitude required to change the contact network.  This implies that although $\delta_c \rightarrow 0$ in the thermodynamic limit, the harmonic approximation should still describe small amplitude perturbations, in complete agreement with our results. The smooth onset of contact nonlinearities for $\alpha>2$ also implies that, provided the time scale of the measurement is suitable, low amplitude \emph{microscopic} measurements, for which $\delta < \delta_0$ , can also be described by linear response. The issue of time scales is important and is discussed in the next section.

\section{Nonzero-amplitude vibrations and time scales\label{sec:finite_amplitude_discussion}}
So far, by considering infinitesimal vibrations, we have shown that linear response can accurately describe the lowest-order behavior of bulk quantities. However, since experiments must study nonzero-amplitude perturbations, it is also important to understand how nonlinearities affect the response.  

First, consider a perturbation in the direction of one of the normal modes of vibration with frequency $\omega_1$. The motion is determined by the position dependent dynamical matrix $D_{ij}(\vec u)$ according to Eqs.~\eqref{eq_of_motion} and \eqref{integral_of_dynamical_matrix}. Ignoring expansion nonlinearities, $D_{ij}(\vec u)$ is constant if contacts do not change and the system undergoes oscillatory motion with a $\delta$-function in the Fourier transform at $\omega_1$. The dynamical matrix changes when a contact forms (or breaks), so when represented by the eigenvectors of the new dynamical matrix the trajectory becomes smeared out over multiple modes. Motion along these modes will evolve at different frequencies and so the system will be in a slightly different position when the contact reopens (or reforms). This leads to mode mixing and a broadening in time of the Fourier transform. 

References~\cite{Schreck:2011kl} and \cite{Schreck:2013gg} showed that for small systems close to jamming, the effect of mode mixing is particularly sudden and dramatic as soon as a contact forms or breaks, {\it i.e.} when $\delta > \delta_c$.
However, mode mixing can also occur without a change to the contact network because expansion nonlinearities cause $D_{ij}(\vec u)$ to change for any $\vec u$.
Indeed, mode mixing is a generic feature of finite-amplitude vibrations even for systems without contact nonlinearities ({\it e.g.} with Lennard-Jones interactions) and for which there is a clear linear regime ({\it e.g.} a crystal)~\cite{Ashcroft:1976ud}. Although mode mixing is ubiquitous, the effect might not be noticeable over short times when $\delta < \delta_c$, as demonstrated by Refs.~\cite{Schreck:2011kl} and \cite{Schreck:2013gg}. 

Therefore, an important factor is the time scale over which a measurement is made. To understand this time scale for a particular system, one must know how the eigenvectors and associated frequencies of $D_{ij}(\vec u)$ change as the system evolves. For example, if the initial trajectory only projects onto nearly degenerate modes, then the broadening of the Fourier transform will be very slow whereas if the trajectory projects onto modes with very different frequencies, then the broadening will be occur quickly. Also of relevance is the amount of time during the oscillation for which $D_{ij}(\vec u)$ differs from $D_{ij}(\vec 0)$.

While this time scale can be important, for example in phonon scattering, it has little relevance to understanding bulk response to leading order, which is the focus of this paper. Importantly, the presence of mode mixing is \emph{not} an indication that there is no linear regime~\cite{Ashcroft:1976ud}. For large amplitude vibrations ($\delta \gg \delta_c$), Refs.~\cite{Schreck:2011kl} and \cite{Schreck:2013gg} showed that the Fourier transform differs greatly from the harmonic prediction. However, it is important to distinguish this from the density of normal modes, which is \emph{defined} from the dynamical matrix and the harmonic approximation. 
We note that while we have focused on the harmonic approximation of the potential energy, one can also think of normal modes of the free energy ({\it e.g.} of hard sphere glasses~\cite{Brito:2009ed}), though they are still defined within the harmonic approximation. 
If one assumes that the harmonic approximation is valid, then there are a variety of ways to calculate the density of states, including also the velocity autocorrelation function and the displacement covariance matrix~\cite{Chen:2010gb,Ghosh:2010hu}. While these approaches are often much more feasible, especially in experimental systems, they only measure the density of states provided the systems remains in the linear regime.

\section{Discussion\label{sec:discussion}}
We have shown that jammed soft sphere packings always have a well-defined linear regime regardless of system size whenever $\Delta \phi > 0$, thus providing sound justification for the use of the harmonic approximation in the study of bulk response. Although Ref.~\cite{Schreck:2011kl} showed that $\delta_c$, which marks the onset of contact nonlinearities, vanishes as $N\rightarrow \infty$, individual contact nonlinearities have a vanishing effect on bulk response in the thermodynamic limit. When measuring microscopic quantities like the evolution over time after a specific perturbation, Schreck {\it et al.}~\cite{Schreck:2011kl,Schreck:2013gg} showed that nonlinear effects are indeed important in jammed packings, just as they are for crystals. Nevertheless, they are \emph{not} essential for understanding bulk response to leading order.

The primary result of this paper is the perturbation argument presented in Sec.~\ref{sec:perturbation_argument}, which is valid for any potential of the form of Eq.~\eqref{pair_potential}. However, note that we only \emph{need} to invoke this argument for the case of Hookian repulsions ($\alpha=2$). The onset of contact nonlinearities is smooth when $\alpha>2$ and thus has the potential to cause problems only when $\alpha \leq 2$. This leads to the interesting and counterintuitive result that nonlinear pair potentials are more harmonic than one-sided linear springs. 

Our results are consistent with the recent work of van Deen {\it et al.}~\cite{vanDeen:2014vm}, who look at jammed sphere packings undergoing quasi-static shear. They measure the ratio of the shear modulus before and after a contact change, which they find to approach unity for $\Delta \phi N^2 \gg 10$~\footnote{This is reported in Ref.~\cite{vanDeen:2014vm} as $pN^2$ not $\Delta \phi N^2$, but these two scalings are equivalent for Hookian interaction, which they use. One would expect the scaling to remain $\Delta \phi N^2$ for $\alpha \neq 2$.}. Note that this scaling was previously shown to control the finite-size behavior of the shear modulus, which only exhibits the canonical $G \sim \Delta \phi^{1/2}$ power law when $\Delta \phi N^2 \gg 10$~\cite{Goodrich:2012ck}. This suggests that if a system is large enough to exhibit this bulk scaling behavior, then it is large enough to be insensitive to individual contact changes. 

When an extensive number of contacts break, van Deen {\it et al.}~\cite{vanDeen:2014vm} also show that fluctuations in the shear modulus scale as $\Delta \phi^{-1/2-2\beta} N^{1-4\beta}$, where $\beta \approx 0.35$. As predicted, the shear modulus converges to a well-defined value in the thermodynamic limit but not in the limit $\Delta \phi \rightarrow 0$. Furthermore, Dagois-Bohy {\it et al.}~\cite{Dagois-Bohy:private:2014} study oscillatory rheology and find that the strain amplitude where linear response breaks down in large systems is independent of system size. This is also consistent with recent simulations by Tighe {\it et al.}~\cite{Tighe2014} that explicitly measure the extent of the linear regime as systems are sheared.

Our results also provide context for the work of Ikeda {\it et al.}~\cite{Ikeda:2013gu}, who studied nonlinearities that arise from thermal fluctuations. They find that nonlinearities begin to modify the linear vibrations when fluctuations in the distance between neighboring particles is comparable to the width of the first peak of the radial distribution function. Such fluctuations cause an extensive number of contacts to break and is therefore in complete agreement with our results.
References~\cite{Ikeda:2013gu} and \cite{Wang:2013gy} show that there is a well defined temperature scale $T^*$ that marks the breakdown of the harmonic approximation. For pair interactions of the form of Eq.~\eqref{pair_potential}, $T^*$ is proportional to $\Delta \phi^\alpha$, though the prefactor depends sensitively on the way one measures nonlinearities~\cite{Ikeda:2013gu,Wang:2013gy}.

Extending this result to experimental systems can be difficult because pair interactions are often not known precisely. 
Nevertheless, our results suggest that, for example, packings of soft colloidal micro-gels at room temperature should display harmonic behavior at high densities but cross over to nonlinear behavior as the density is lowered. In a recent experiment, Still {\it et al.}~\cite{Still:2014bd} measured the elastic moduli of a PNIPAM glass by calculating the dispersion relation from the displacement covariance matrix. They found clean harmonic behavior over a range of densities that agrees nicely with numerical calculations~\cite{Shundyak:2007ga,Somfai:2007ge} of frictional soft spheres.

Although expansion nonlinearities guarantee that the density of normal modes differs from the infinite-time spectral density of finite-amplitude vibrations, the harmonic approximation nonetheless provides the foundation from which we can understand such nonlinear behavior. Indeed, many aspects of nonlinear response are strongly correlated with linear-response properties.  For example, the Gruneisen parameter, an anharmonic property, depends on mode frequency, a harmonic property, in a way that is understood~\cite{Xu:2010fa}. The energy barrier to rearrangement in a given mode direction is strongly correlated with mode frequency as well~\cite{Xu:2010fa}, and the spatial location of particle rearrangements is strongly correlated with high-displacement regions in quasi-localized low-frequency modes~\cite{WidmerCooper:2009dp,Tanguy:2010dh,Manning:2011dk}.

Even at the onset of jamming, where the linear regime vanishes, it is essential to understand the linear response in order to approach the nonlinear response. This is illustrated by a recent analysis of shock waves in marginally jammed solids~\cite{Gomez:2012ji}.  The importance of linear quantities in the presence of a vanishingly small linear regime is not unique to jamming. 
In the Ising model, for example, the magnetic susceptibility diverges at the critical point, but the linear theory is still central to our understanding of the phase transition. Just as one must first understand phonons to understand phonon-phonon scattering, the density of normal modes and other linear response properties provide essential insight into the nature of jammed solids. 

We thank Justin Burton, Lisa Manning, Samuel Schoenholz, Anton Souslov, Daniel Sussman, Brian Tighe and Martin van Hecke for important discussions.
This research was supported by the U.S. Department of Energy, Office of Basic Energy Sciences, Division of Materials Sciences and Engineering under Award DE-FG02-05ER46199 (AJL and CPG) and DE-FG02-03ER46088 (SN). CPG was partially supported by the NSF through a Graduate Research Fellowship.


\begin{thebibliography}{36}%
\makeatletter
\providecommand \@ifxundefined [1]{%
 \@ifx{#1\undefined}
}%
\providecommand \@ifnum [1]{%
 \ifnum #1\expandafter \@firstoftwo
 \else \expandafter \@secondoftwo
 \fi
}%
\providecommand \@ifx [1]{%
 \ifx #1\expandafter \@firstoftwo
 \else \expandafter \@secondoftwo
 \fi
}%
\providecommand \natexlab [1]{#1}%
\providecommand \enquote  [1]{``#1''}%
\providecommand \bibnamefont  [1]{#1}%
\providecommand \bibfnamefont [1]{#1}%
\providecommand \citenamefont [1]{#1}%
\providecommand \href@noop [0]{\@secondoftwo}%
\providecommand \href [0]{\begingroup \@sanitize@url \@href}%
\providecommand \@href[1]{\@@startlink{#1}\@@href}%
\providecommand \@@href[1]{\endgroup#1\@@endlink}%
\providecommand \@sanitize@url [0]{\catcode `\\12\catcode `\$12\catcode
  `\&12\catcode `\#12\catcode `\^12\catcode `\_12\catcode `\%12\relax}%
\providecommand \@@startlink[1]{}%
\providecommand \@@endlink[0]{}%
\providecommand \url  [0]{\begingroup\@sanitize@url \@url }%
\providecommand \@url [1]{\endgroup\@href {#1}{\urlprefix }}%
\providecommand \urlprefix  [0]{URL }%
\providecommand \Eprint [0]{\href }%
\providecommand \doibase [0]{http://dx.doi.org/}%
\providecommand \selectlanguage [0]{\@gobble}%
\providecommand \bibinfo  [0]{\@secondoftwo}%
\providecommand \bibfield  [0]{\@secondoftwo}%
\providecommand \translation [1]{[#1]}%
\providecommand \BibitemOpen [0]{}%
\providecommand \bibitemStop [0]{}%
\providecommand \bibitemNoStop [0]{.\EOS\space}%
\providecommand \EOS [0]{\spacefactor3000\relax}%
\providecommand \BibitemShut  [1]{\csname bibitem#1\endcsname}%
\let\auto@bib@innerbib\@empty
\bibitem [{\citenamefont {Ashcroft}\ and\ \citenamefont
  {Mermin}(1976{\natexlab{a}})}]{Ashcroft:1976ud}%
  \BibitemOpen
  \bibfield  {author} {\bibinfo {author} {\bibfnamefont {N.~W.}\ \bibnamefont
  {Ashcroft}}\ and\ \bibinfo {author} {\bibfnamefont {N.~D.}\ \bibnamefont
  {Mermin}},\ }\href@noop {} {\emph {\bibinfo {title} {{Solid state
  physics}}}}\ (\bibinfo  {publisher} {Thomson Brooks/Cole},\ \bibinfo {year}
  {1976})\BibitemShut {NoStop}%
\bibitem [{\citenamefont {Landau}\ \emph {et~al.}(1986)\citenamefont {Landau},
  \citenamefont {Pitaevskii}, \citenamefont {Lifshitz},\ and\ \citenamefont
  {Kosevich}}]{Landau1986}%
  \BibitemOpen
  \bibfield  {author} {\bibinfo {author} {\bibfnamefont {L.~D.}\ \bibnamefont
  {Landau}}, \bibinfo {author} {\bibfnamefont {L.~P.}\ \bibnamefont
  {Pitaevskii}}, \bibinfo {author} {\bibfnamefont {E.~M.}\ \bibnamefont
  {Lifshitz}}, \ and\ \bibinfo {author} {\bibfnamefont {A.~M.}\ \bibnamefont
  {Kosevich}},\ }\href@noop {} {\emph {\bibinfo {title} {Theory of
  Elasticity}}},\ \bibinfo {edition} {3rd}\ ed.\ (\bibinfo  {publisher}
  {Butterworth-Heinemann},\ \bibinfo {year} {1986})\BibitemShut {NoStop}%
\bibitem [{\citenamefont {Liu}\ and\ \citenamefont {Nagel}(2010)}]{Liu:2010jx}%
  \BibitemOpen
  \bibfield  {author} {\bibinfo {author} {\bibfnamefont {A.~J.}\ \bibnamefont
  {Liu}}\ and\ \bibinfo {author} {\bibfnamefont {S.~R.}\ \bibnamefont
  {Nagel}},\ }\href@noop {} {\bibfield  {journal} {\bibinfo  {journal} {Annu.
  Rev. Condens. Matter Phys.}\ }\textbf {\bibinfo {volume} {1}},\ \bibinfo
  {pages} {347} (\bibinfo {year} {2010})}\BibitemShut {NoStop}%
\bibitem [{\citenamefont {Liu}\ and\ \citenamefont {Nagel}(1998)}]{Liu:1998up}%
  \BibitemOpen
  \bibfield  {author} {\bibinfo {author} {\bibfnamefont {A.~J.}\ \bibnamefont
  {Liu}}\ and\ \bibinfo {author} {\bibfnamefont {S.~R.}\ \bibnamefont
  {Nagel}},\ }\href@noop {} {\bibfield  {journal} {\bibinfo  {journal}
  {Nature}\ }\textbf {\bibinfo {volume} {396}},\ \bibinfo {pages} {21}
  (\bibinfo {year} {1998})}\BibitemShut {NoStop}%
\bibitem [{\citenamefont {O'Hern}\ \emph {et~al.}(2003)\citenamefont {O'Hern},
  \citenamefont {Silbert}, \citenamefont {Liu},\ and\ \citenamefont
  {Nagel}}]{OHern:2003vq}%
  \BibitemOpen
  \bibfield  {author} {\bibinfo {author} {\bibfnamefont {C.~S.}\ \bibnamefont
  {O'Hern}}, \bibinfo {author} {\bibfnamefont {L.~E.}\ \bibnamefont {Silbert}},
  \bibinfo {author} {\bibfnamefont {A.~J.}\ \bibnamefont {Liu}}, \ and\
  \bibinfo {author} {\bibfnamefont {S.~R.}\ \bibnamefont {Nagel}},\ }\href@noop
  {} {\bibfield  {journal} {\bibinfo  {journal} {Phys. Rev. E}\ }\textbf
  {\bibinfo {volume} {68}},\ \bibinfo {pages} {011306} (\bibinfo {year}
  {2003})}\BibitemShut {NoStop}%
\bibitem [{\citenamefont {Wyart}\ \emph {et~al.}(2005)\citenamefont {Wyart},
  \citenamefont {Nagel},\ and\ \citenamefont {Witten}}]{Wyart:2005wv}%
  \BibitemOpen
  \bibfield  {author} {\bibinfo {author} {\bibfnamefont {M.}~\bibnamefont
  {Wyart}}, \bibinfo {author} {\bibfnamefont {S.~R.}\ \bibnamefont {Nagel}}, \
  and\ \bibinfo {author} {\bibfnamefont {T.~A.}\ \bibnamefont {Witten}},\
  }\href@noop {} {\bibfield  {journal} {\bibinfo  {journal} {EPL}\ }\textbf
  {\bibinfo {volume} {72}},\ \bibinfo {pages} {486} (\bibinfo {year}
  {2005})}\BibitemShut {NoStop}%
\bibitem [{\citenamefont {Silbert}\ \emph {et~al.}(2005)\citenamefont
  {Silbert}, \citenamefont {Liu},\ and\ \citenamefont
  {Nagel}}]{Silbert:2005vw}%
  \BibitemOpen
  \bibfield  {author} {\bibinfo {author} {\bibfnamefont {L.~E.}\ \bibnamefont
  {Silbert}}, \bibinfo {author} {\bibfnamefont {A.~J.}\ \bibnamefont {Liu}}, \
  and\ \bibinfo {author} {\bibfnamefont {S.~R.}\ \bibnamefont {Nagel}},\
  }\href@noop {} {\bibfield  {journal} {\bibinfo  {journal} {Phys. Rev. Lett.}\
  }\textbf {\bibinfo {volume} {95}},\ \bibinfo {pages} {098301} (\bibinfo
  {year} {2005})}\BibitemShut {NoStop}%
\bibitem [{\citenamefont {Ellenbroek}\ \emph {et~al.}(2006)\citenamefont
  {Ellenbroek}, \citenamefont {Somfai}, \citenamefont {van Hecke},\ and\
  \citenamefont {van Saarloos}}]{Ellenbroek:2006df}%
  \BibitemOpen
  \bibfield  {author} {\bibinfo {author} {\bibfnamefont {W.~G.}\ \bibnamefont
  {Ellenbroek}}, \bibinfo {author} {\bibfnamefont {E.}~\bibnamefont {Somfai}},
  \bibinfo {author} {\bibfnamefont {M.}~\bibnamefont {van Hecke}}, \ and\
  \bibinfo {author} {\bibfnamefont {W.}~\bibnamefont {van Saarloos}},\
  }\href@noop {} {\bibfield  {journal} {\bibinfo  {journal} {Phys. Rev. Lett.}\
  }\textbf {\bibinfo {volume} {97}},\ \bibinfo {pages} {258001} (\bibinfo
  {year} {2006})}\BibitemShut {NoStop}%
\bibitem [{\citenamefont {Xu}\ \emph {et~al.}(2007)\citenamefont {Xu},
  \citenamefont {Wyart}, \citenamefont {Liu},\ and\ \citenamefont
  {Nagel}}]{Xu:2007dd}%
  \BibitemOpen
  \bibfield  {author} {\bibinfo {author} {\bibfnamefont {N.}~\bibnamefont
  {Xu}}, \bibinfo {author} {\bibfnamefont {M.}~\bibnamefont {Wyart}}, \bibinfo
  {author} {\bibfnamefont {A.~J.}\ \bibnamefont {Liu}}, \ and\ \bibinfo
  {author} {\bibfnamefont {S.~R.}\ \bibnamefont {Nagel}},\ }\href@noop {}
  {\bibfield  {journal} {\bibinfo  {journal} {Phys. Rev. Lett.}\ }\textbf
  {\bibinfo {volume} {98}},\ \bibinfo {pages} {175502} (\bibinfo {year}
  {2007})}\BibitemShut {NoStop}%
\bibitem [{\citenamefont {Schreck}\ \emph {et~al.}(2010)\citenamefont
  {Schreck}, \citenamefont {Xu},\ and\ \citenamefont {O'Hern}}]{Xu:2010vd}%
  \BibitemOpen
  \bibfield  {author} {\bibinfo {author} {\bibfnamefont {C.~F.}\ \bibnamefont
  {Schreck}}, \bibinfo {author} {\bibfnamefont {N.}~\bibnamefont {Xu}}, \ and\
  \bibinfo {author} {\bibfnamefont {C.~S.}\ \bibnamefont {O'Hern}},\
  }\href@noop {} {\bibfield  {journal} {\bibinfo  {journal} {Soft Matter}\
  }\textbf {\bibinfo {volume} {6}},\ \bibinfo {pages} {2960} (\bibinfo {year}
  {2010})}\BibitemShut {NoStop}%
\bibitem [{\citenamefont {Goodrich}\ \emph {et~al.}(2012)\citenamefont
  {Goodrich}, \citenamefont {Liu},\ and\ \citenamefont
  {Nagel}}]{Goodrich:2012ck}%
  \BibitemOpen
  \bibfield  {author} {\bibinfo {author} {\bibfnamefont {C.~P.}\ \bibnamefont
  {Goodrich}}, \bibinfo {author} {\bibfnamefont {A.~J.}\ \bibnamefont {Liu}}, \
  and\ \bibinfo {author} {\bibfnamefont {S.~R.}\ \bibnamefont {Nagel}},\
  }\href@noop {} {\bibfield  {journal} {\bibinfo  {journal} {Phys. Rev. Lett.}\
  }\textbf {\bibinfo {volume} {109}},\ \bibinfo {pages} {095704} (\bibinfo
  {year} {2012})}\BibitemShut {NoStop}%
\bibitem [{\citenamefont {Schreck}\ \emph {et~al.}(2011)\citenamefont
  {Schreck}, \citenamefont {Bertrand}, \citenamefont {O'Hern},\ and\
  \citenamefont {Shattuck}}]{Schreck:2011kl}%
  \BibitemOpen
  \bibfield  {author} {\bibinfo {author} {\bibfnamefont {C.~F.}\ \bibnamefont
  {Schreck}}, \bibinfo {author} {\bibfnamefont {T.}~\bibnamefont {Bertrand}},
  \bibinfo {author} {\bibfnamefont {C.~S.}\ \bibnamefont {O'Hern}}, \ and\
  \bibinfo {author} {\bibfnamefont {M.~D.}\ \bibnamefont {Shattuck}},\
  }\href@noop {} {\bibfield  {journal} {\bibinfo  {journal} {Phys. Rev. Lett.}\
  }\textbf {\bibinfo {volume} {107}},\ \bibinfo {pages} {078301} (\bibinfo
  {year} {2011})}\BibitemShut {NoStop}%
\bibitem [{\citenamefont {Hentschel}\ \emph {et~al.}(2011)\citenamefont
  {Hentschel}, \citenamefont {Karmakar}, \citenamefont {Lerner},\ and\
  \citenamefont {Procaccia}}]{Hentschel:2011cba}%
  \BibitemOpen
  \bibfield  {author} {\bibinfo {author} {\bibfnamefont {H.~G.~E.}\
  \bibnamefont {Hentschel}}, \bibinfo {author} {\bibfnamefont {S.}~\bibnamefont
  {Karmakar}}, \bibinfo {author} {\bibfnamefont {E.}~\bibnamefont {Lerner}}, \
  and\ \bibinfo {author} {\bibfnamefont {I.}~\bibnamefont {Procaccia}},\
  }\href@noop {} {\bibfield  {journal} {\bibinfo  {journal} {Phys. Rev. E}\
  }\textbf {\bibinfo {volume} {83}},\ \bibinfo {pages} {061101} (\bibinfo
  {year} {2011})}\BibitemShut {NoStop}%
\bibitem [{\citenamefont {Goodrich}\ \emph {et~al.}(2014)\citenamefont
  {Goodrich}, \citenamefont {Liu},\ and\ \citenamefont
  {Nagel}}]{Goodrich:2014jw}%
  \BibitemOpen
  \bibfield  {author} {\bibinfo {author} {\bibfnamefont {C.~P.}\ \bibnamefont
  {Goodrich}}, \bibinfo {author} {\bibfnamefont {A.~J.}\ \bibnamefont {Liu}}, \
  and\ \bibinfo {author} {\bibfnamefont {S.~R.}\ \bibnamefont {Nagel}},\
  }\href@noop {} {\bibfield  {journal} {\bibinfo  {journal} {Phys. Rev. Lett.}\
  }\textbf {\bibinfo {volume} {112}},\ \bibinfo {pages} {049801} (\bibinfo
  {year} {2014})}\BibitemShut {NoStop}%
\bibitem [{Note1()}]{Note1}%
  \BibitemOpen
  \bibinfo {note} {It is not necessary for $\protect \genfrac {}{}{}112
  D^0_{ij}u_iu_j$ to be positive in the directions of global translation
  because the energy landscape in these directions is completely
  flat.}\BibitemShut {Stop}%
\bibitem [{Note2()}]{Note2}%
  \BibitemOpen
  \bibinfo {note} {A notable exception is packings of particles with one-sided
  Hookian interactions.}\BibitemShut {Stop}%
\bibitem [{\citenamefont {Ashcroft}\ and\ \citenamefont
  {Mermin}(1976{\natexlab{b}})}]{Ashcroft1976}%
  \BibitemOpen
  \bibfield  {author} {\bibinfo {author} {\bibfnamefont {N.~W.}\ \bibnamefont
  {Ashcroft}}\ and\ \bibinfo {author} {\bibfnamefont {N.~D.}\ \bibnamefont
  {Mermin}},\ }\href@noop {} {\emph {\bibinfo {title} {Solid State Physics}}},\
  \bibinfo {edition} {1st}\ ed.\ (\bibinfo  {publisher} {Thomson Learning},\
  \bibinfo {address} {Toronto},\ \bibinfo {year} {1976})\BibitemShut {NoStop}%
\bibitem [{\citenamefont {Xu}\ \emph {et~al.}(2010)\citenamefont {Xu},
  \citenamefont {Vitelli}, \citenamefont {Liu},\ and\ \citenamefont
  {Nagel}}]{Xu:2010fa}%
  \BibitemOpen
  \bibfield  {author} {\bibinfo {author} {\bibfnamefont {N.}~\bibnamefont
  {Xu}}, \bibinfo {author} {\bibfnamefont {V.}~\bibnamefont {Vitelli}},
  \bibinfo {author} {\bibfnamefont {A.~J.}\ \bibnamefont {Liu}}, \ and\
  \bibinfo {author} {\bibfnamefont {S.~R.}\ \bibnamefont {Nagel}},\ }\href@noop
  {} {\bibfield  {journal} {\bibinfo  {journal} {EPL}\ }\textbf {\bibinfo
  {volume} {90}},\ \bibinfo {pages} {56001} (\bibinfo {year}
  {2010})}\BibitemShut {NoStop}%
\bibitem [{\citenamefont {G{\'o}mez}\ \emph {et~al.}(2012)\citenamefont
  {G{\'o}mez}, \citenamefont {Turner}, \citenamefont {van Hecke},\ and\
  \citenamefont {Vitelli}}]{Gomez:2012ji}%
  \BibitemOpen
  \bibfield  {author} {\bibinfo {author} {\bibfnamefont {L.~R.}\ \bibnamefont
  {G{\'o}mez}}, \bibinfo {author} {\bibfnamefont {A.~M.}\ \bibnamefont
  {Turner}}, \bibinfo {author} {\bibfnamefont {M.}~\bibnamefont {van Hecke}}, \
  and\ \bibinfo {author} {\bibfnamefont {V.}~\bibnamefont {Vitelli}},\
  }\href@noop {} {\bibfield  {journal} {\bibinfo  {journal} {Phys. Rev. Lett.}\
  }\textbf {\bibinfo {volume} {108}},\ \bibinfo {pages} {058001} (\bibinfo
  {year} {2012})}\BibitemShut {NoStop}%
\bibitem [{\citenamefont {Ikeda}\ \emph {et~al.}(2013)\citenamefont {Ikeda},
  \citenamefont {Berthier},\ and\ \citenamefont {Biroli}}]{Ikeda:2013gu}%
  \BibitemOpen
  \bibfield  {author} {\bibinfo {author} {\bibfnamefont {A.}~\bibnamefont
  {Ikeda}}, \bibinfo {author} {\bibfnamefont {L.}~\bibnamefont {Berthier}}, \
  and\ \bibinfo {author} {\bibfnamefont {G.}~\bibnamefont {Biroli}},\
  }\href@noop {} {\bibfield  {journal} {\bibinfo  {journal} {J. Chem. Phys.}\
  }\textbf {\bibinfo {volume} {138}},\ \bibinfo {pages} {12A507} (\bibinfo
  {year} {2013})}\BibitemShut {NoStop}%
\bibitem [{\citenamefont {Schreck}\ \emph {et~al.}(2013)\citenamefont
  {Schreck}, \citenamefont {O'Hern},\ and\ \citenamefont
  {Shattuck}}]{Schreck:2013gg}%
  \BibitemOpen
  \bibfield  {author} {\bibinfo {author} {\bibfnamefont {C.~F.}\ \bibnamefont
  {Schreck}}, \bibinfo {author} {\bibfnamefont {C.~S.}\ \bibnamefont {O'Hern}},
  \ and\ \bibinfo {author} {\bibfnamefont {M.~D.}\ \bibnamefont {Shattuck}},\
  }\href@noop {} {\bibfield  {journal} {\bibinfo  {journal} {Granular Matter}\
  } (\bibinfo {year} {2013})}\BibitemShut {NoStop}%
\bibitem [{\citenamefont {Wang}\ and\ \citenamefont {Xu}(2013)}]{Wang:2013gy}%
  \BibitemOpen
  \bibfield  {author} {\bibinfo {author} {\bibfnamefont {L.}~\bibnamefont
  {Wang}}\ and\ \bibinfo {author} {\bibfnamefont {N.}~\bibnamefont {Xu}},\
  }\href@noop {} {\bibfield  {journal} {\bibinfo  {journal} {Soft Matter}\
  }\textbf {\bibinfo {volume} {9}},\ \bibinfo {pages} {2475} (\bibinfo {year}
  {2013})}\BibitemShut {NoStop}%
\bibitem [{\citenamefont {Pellegrino}(1993)}]{Pellegrino:1993uu}%
  \BibitemOpen
  \bibfield  {author} {\bibinfo {author} {\bibfnamefont {S.}~\bibnamefont
  {Pellegrino}},\ }\href@noop {} {\bibfield  {journal} {\bibinfo  {journal}
  {International Journal of Solids and Structures}\ }\textbf {\bibinfo {volume}
  {30}},\ \bibinfo {pages} {3025} (\bibinfo {year} {1993})}\BibitemShut
  {NoStop}%
\bibitem [{\citenamefont {Brito}\ and\ \citenamefont
  {Wyart}(2009)}]{Brito:2009ed}%
  \BibitemOpen
  \bibfield  {author} {\bibinfo {author} {\bibfnamefont {C.}~\bibnamefont
  {Brito}}\ and\ \bibinfo {author} {\bibfnamefont {M.}~\bibnamefont {Wyart}},\
  }\href@noop {} {\bibfield  {journal} {\bibinfo  {journal} {J. Chem. Phys.}\
  }\textbf {\bibinfo {volume} {131}},\ \bibinfo {pages} {024504} (\bibinfo
  {year} {2009})}\BibitemShut {NoStop}%
\bibitem [{\citenamefont {Chen}\ \emph {et~al.}(2010)\citenamefont {Chen},
  \citenamefont {Ellenbroek}, \citenamefont {Zhang}, \citenamefont {Chen},
  \citenamefont {Yunker}, \citenamefont {Henkes}, \citenamefont {Brito},
  \citenamefont {Dauchot}, \citenamefont {van Saarloos}, \citenamefont {Liu},\
  and\ \citenamefont {Yodh}}]{Chen:2010gb}%
  \BibitemOpen
  \bibfield  {author} {\bibinfo {author} {\bibfnamefont {K.}~\bibnamefont
  {Chen}}, \bibinfo {author} {\bibfnamefont {W.~G.}\ \bibnamefont
  {Ellenbroek}}, \bibinfo {author} {\bibfnamefont {Z.}~\bibnamefont {Zhang}},
  \bibinfo {author} {\bibfnamefont {D.~T.~N.}\ \bibnamefont {Chen}}, \bibinfo
  {author} {\bibfnamefont {P.~J.}\ \bibnamefont {Yunker}}, \bibinfo {author}
  {\bibfnamefont {S.}~\bibnamefont {Henkes}}, \bibinfo {author} {\bibfnamefont
  {C.}~\bibnamefont {Brito}}, \bibinfo {author} {\bibfnamefont
  {O.}~\bibnamefont {Dauchot}}, \bibinfo {author} {\bibfnamefont
  {W.}~\bibnamefont {van Saarloos}}, \bibinfo {author} {\bibfnamefont {A.~J.}\
  \bibnamefont {Liu}}, \ and\ \bibinfo {author} {\bibfnamefont {A.~G.}\
  \bibnamefont {Yodh}},\ }\href@noop {} {\bibfield  {journal} {\bibinfo
  {journal} {Phys. Rev. Lett.}\ }\textbf {\bibinfo {volume} {105}},\ \bibinfo
  {pages} {025501} (\bibinfo {year} {2010})}\BibitemShut {NoStop}%
\bibitem [{\citenamefont {Ghosh}\ \emph {et~al.}(2010)\citenamefont {Ghosh},
  \citenamefont {Chikkadi}, \citenamefont {Schall}, \citenamefont {Kurchan},\
  and\ \citenamefont {Bonn}}]{Ghosh:2010hu}%
  \BibitemOpen
  \bibfield  {author} {\bibinfo {author} {\bibfnamefont {A.}~\bibnamefont
  {Ghosh}}, \bibinfo {author} {\bibfnamefont {V.~K.}\ \bibnamefont {Chikkadi}},
  \bibinfo {author} {\bibfnamefont {P.}~\bibnamefont {Schall}}, \bibinfo
  {author} {\bibfnamefont {J.}~\bibnamefont {Kurchan}}, \ and\ \bibinfo
  {author} {\bibfnamefont {D.}~\bibnamefont {Bonn}},\ }\href@noop {} {\bibfield
   {journal} {\bibinfo  {journal} {Phys. Rev. Lett.}\ }\textbf {\bibinfo
  {volume} {104}},\ \bibinfo {pages} {248305} (\bibinfo {year}
  {2010})}\BibitemShut {NoStop}%
\bibitem [{\citenamefont {van Deen}\ \emph {et~al.}(2014)\citenamefont {van
  Deen}, \citenamefont {Simon}, \citenamefont {Zeravcic}, \citenamefont
  {Dagois-Bohy}, \citenamefont {Tighe},\ and\ \citenamefont {van
  Hecke}}]{vanDeen:2014vm}%
  \BibitemOpen
  \bibfield  {author} {\bibinfo {author} {\bibfnamefont {M.~S.}\ \bibnamefont
  {van Deen}}, \bibinfo {author} {\bibfnamefont {J.}~\bibnamefont {Simon}},
  \bibinfo {author} {\bibfnamefont {Z.}~\bibnamefont {Zeravcic}}, \bibinfo
  {author} {\bibfnamefont {S.}~\bibnamefont {Dagois-Bohy}}, \bibinfo {author}
  {\bibfnamefont {B.~P.}\ \bibnamefont {Tighe}}, \ and\ \bibinfo {author}
  {\bibfnamefont {M.}~\bibnamefont {van Hecke}},\ }\href@noop {} {\bibfield
  {journal} {\bibinfo  {journal} {arXiv}\ } (\bibinfo {year} {2014})},\ \Eprint
  {http://arxiv.org/abs/1404.3156v1} {1404.3156v1} \BibitemShut {NoStop}%
\bibitem [{Note3()}]{Note3}%
  \BibitemOpen
  \bibinfo {note} {This is reported in Ref.~\cite {vanDeen:2014vm} as $pN^2$
  not $\Delta \phi N^2$, but these two scalings are equivalent for Hookian
  interaction, which they use. One would expect the scaling to remain $\Delta
  \phi N^2$ for $\alpha \not =2$.}\BibitemShut {Stop}%
\bibitem [{\citenamefont {Dagois-Bohy}\ \emph {et~al.}(2014)\citenamefont
  {Dagois-Bohy}, \citenamefont {Somfai}, \citenamefont {Tighe},\ and\
  \citenamefont {van Hecke}}]{Dagois-Bohy:private:2014}%
  \BibitemOpen
  \bibfield  {author} {\bibinfo {author} {\bibfnamefont {S.}~\bibnamefont
  {Dagois-Bohy}}, \bibinfo {author} {\bibfnamefont {E.}~\bibnamefont {Somfai}},
  \bibinfo {author} {\bibfnamefont {B.~P.}\ \bibnamefont {Tighe}}, \ and\
  \bibinfo {author} {\bibfnamefont {M.}~\bibnamefont {van Hecke}},\ }\href@noop
  {} {} (\bibinfo {year} {2014}),\ \bibinfo {note} {preprint}\BibitemShut
  {NoStop}%
\bibitem [{\citenamefont {Tighe}\ \emph {et~al.}(2014)\citenamefont {Tighe},
  \citenamefont {Boschan},\ and\ \citenamefont {Somfai}}]{Tighe2014}%
  \BibitemOpen
  \bibfield  {author} {\bibinfo {author} {\bibfnamefont {B.}~\bibnamefont
  {Tighe}}, \bibinfo {author} {\bibfnamefont {J.}~\bibnamefont {Boschan}}, \
  and\ \bibinfo {author} {\bibfnamefont {E.}~\bibnamefont {Somfai}}\ }(\bibinfo
   {publisher} {APS March Meeting},\ \bibinfo {year} {2014})\BibitemShut
  {NoStop}%
\bibitem [{\citenamefont {Still}\ \emph {et~al.}(2014)\citenamefont {Still},
  \citenamefont {Goodrich}, \citenamefont {Chen}, \citenamefont {Yunker},
  \citenamefont {Schoenholz}, \citenamefont {Liu},\ and\ \citenamefont
  {Yodh}}]{Still:2014bd}%
  \BibitemOpen
  \bibfield  {author} {\bibinfo {author} {\bibfnamefont {T.}~\bibnamefont
  {Still}}, \bibinfo {author} {\bibfnamefont {C.~P.}\ \bibnamefont {Goodrich}},
  \bibinfo {author} {\bibfnamefont {K.}~\bibnamefont {Chen}}, \bibinfo {author}
  {\bibfnamefont {P.~J.}\ \bibnamefont {Yunker}}, \bibinfo {author}
  {\bibfnamefont {S.}~\bibnamefont {Schoenholz}}, \bibinfo {author}
  {\bibfnamefont {A.~J.}\ \bibnamefont {Liu}}, \ and\ \bibinfo {author}
  {\bibfnamefont {A.~G.}\ \bibnamefont {Yodh}},\ }\href@noop {} {\bibfield
  {journal} {\bibinfo  {journal} {Phys. Rev. E}\ }\textbf {\bibinfo {volume}
  {89}},\ \bibinfo {pages} {012301} (\bibinfo {year} {2014})}\BibitemShut
  {NoStop}%
\bibitem [{\citenamefont {Shundyak}\ \emph {et~al.}(2007)\citenamefont
  {Shundyak}, \citenamefont {van Hecke},\ and\ \citenamefont {van
  Saarloos}}]{Shundyak:2007ga}%
  \BibitemOpen
  \bibfield  {author} {\bibinfo {author} {\bibfnamefont {K.}~\bibnamefont
  {Shundyak}}, \bibinfo {author} {\bibfnamefont {M.}~\bibnamefont {van Hecke}},
  \ and\ \bibinfo {author} {\bibfnamefont {W.}~\bibnamefont {van Saarloos}},\
  }\href@noop {} {\bibfield  {journal} {\bibinfo  {journal} {Phys. Rev. E}\
  }\textbf {\bibinfo {volume} {75}},\ \bibinfo {pages} {010301} (\bibinfo
  {year} {2007})}\BibitemShut {NoStop}%
\bibitem [{\citenamefont {Somfai}\ \emph {et~al.}(2007)\citenamefont {Somfai},
  \citenamefont {van Hecke}, \citenamefont {Ellenbroek}, \citenamefont
  {Shundyak},\ and\ \citenamefont {van Saarloos}}]{Somfai:2007ge}%
  \BibitemOpen
  \bibfield  {author} {\bibinfo {author} {\bibfnamefont {E.}~\bibnamefont
  {Somfai}}, \bibinfo {author} {\bibfnamefont {M.}~\bibnamefont {van Hecke}},
  \bibinfo {author} {\bibfnamefont {W.~G.}\ \bibnamefont {Ellenbroek}},
  \bibinfo {author} {\bibfnamefont {K.}~\bibnamefont {Shundyak}}, \ and\
  \bibinfo {author} {\bibfnamefont {W.}~\bibnamefont {van Saarloos}},\
  }\href@noop {} {\bibfield  {journal} {\bibinfo  {journal} {Phys Rev E}\
  }\textbf {\bibinfo {volume} {75}},\ \bibinfo {pages} {020301} (\bibinfo
  {year} {2007})}\BibitemShut {NoStop}%
\bibitem [{\citenamefont {Widmer-Cooper}\ \emph {et~al.}(2009)\citenamefont
  {Widmer-Cooper}, \citenamefont {Perry}, \citenamefont {Harrowell},\ and\
  \citenamefont {Reichman}}]{WidmerCooper:2009dp}%
  \BibitemOpen
  \bibfield  {author} {\bibinfo {author} {\bibfnamefont {A.}~\bibnamefont
  {Widmer-Cooper}}, \bibinfo {author} {\bibfnamefont {H.}~\bibnamefont
  {Perry}}, \bibinfo {author} {\bibfnamefont {P.}~\bibnamefont {Harrowell}}, \
  and\ \bibinfo {author} {\bibfnamefont {D.~R.}\ \bibnamefont {Reichman}},\
  }\href@noop {} {\bibfield  {journal} {\bibinfo  {journal} {J. Chem. Phys.}\
  }\textbf {\bibinfo {volume} {131}},\ \bibinfo {pages} {194508} (\bibinfo
  {year} {2009})}\BibitemShut {NoStop}%
\bibitem [{\citenamefont {Tanguy}\ \emph {et~al.}(2010)\citenamefont {Tanguy},
  \citenamefont {Mantisi},\ and\ \citenamefont {Tsamados}}]{Tanguy:2010dh}%
  \BibitemOpen
  \bibfield  {author} {\bibinfo {author} {\bibfnamefont {A.}~\bibnamefont
  {Tanguy}}, \bibinfo {author} {\bibfnamefont {B.}~\bibnamefont {Mantisi}}, \
  and\ \bibinfo {author} {\bibfnamefont {M.}~\bibnamefont {Tsamados}},\
  }\href@noop {} {\bibfield  {journal} {\bibinfo  {journal} {EPL}\ }\textbf
  {\bibinfo {volume} {90}},\ \bibinfo {pages} {16004} (\bibinfo {year}
  {2010})}\BibitemShut {NoStop}%
\bibitem [{\citenamefont {Manning}\ and\ \citenamefont
  {Liu}(2011)}]{Manning:2011dk}%
  \BibitemOpen
  \bibfield  {author} {\bibinfo {author} {\bibfnamefont {M.~L.}\ \bibnamefont
  {Manning}}\ and\ \bibinfo {author} {\bibfnamefont {A.~J.}\ \bibnamefont
  {Liu}},\ }\href@noop {} {\bibfield  {journal} {\bibinfo  {journal} {Phys.
  Rev. Lett.}\ }\textbf {\bibinfo {volume} {107}},\ \bibinfo {pages} {108302}
  (\bibinfo {year} {2011})}\BibitemShut {NoStop}%
\end{thebibliography}
\end{document}